\documentclass[10pt,journal]{IEEEtran}
\usepackage{amsmath,amsfonts}
\usepackage{algorithmicx}
\usepackage{algorithm}
\usepackage{array}
\usepackage[caption=false,font=normalsize,labelfont=sf,textfont=sf]{subfig}
\usepackage{textcomp}
\usepackage{stfloats}
\usepackage{url}
\usepackage{verbatim}
\usepackage{graphicx}
\usepackage{cite}
\usepackage{multirow}
\usepackage{color}
\usepackage{booktabs}
\usepackage{xcolor}
\begin{document}

\title{Phase Shift Compression for Control Signaling Reduction in IRS-Aided Wireless Systems: Global Attention and Lightweight Design}

\author{Xianhua Yu, Dong Li,~\IEEEmembership{Senior Member,~IEEE}
\thanks{Xianhua Yu and Dong Li are with the School of Computer Science and Engineering, Macau University of Science and Technology, Macau 999078, China (e-mails: 2009853gii30014@student.must.edu.mo; dli@must.edu.mo).}
}



\maketitle
\thispagestyle{empty}
\pagestyle{empty}

\begin{abstract}
A potential 6G technology known as intelligent reflecting surface (IRS) has recently gained much attention from academia and industry. However, acquiring the optimized quantized phase shift (QPS) presents challenges for the IRS due to the phenomenon of signaling storms. In this paper, we attempt to solve the above problem by proposing two deep learning models, the global attention phase shift compression network (GAPSCN) and the simplified GAPSCN (S-GAPSCN). In GAPSCN, we propose a novel attention mechanism that emphasizes a greater number of meaningful features than previous attention-related works. Additionally, S-GAPSCN is built with an asymmetric architecture to meet the practical constraints on computation resources of the IRS controller. Moreover, in S-GAPSCN, to compensate for the performance degradation caused by simplifying the model, we design a low-computation complexity joint attention-assisted multi-scale network (JAAMSN) module in the decoder of S-GAPSCN. Simulation results demonstrate that the proposed global attention mechanism achieves prominent performance compared with the existing attention mechanisms and the proposed GAPSCN can achieve reliable reconstruction performance compared with existing state-of-the-art models. Furthermore, the proposed S-GAPSCN can approach the performance of the GAPSCN at a much lower computational cost.
\end{abstract}

\color{black}


\section{Introduction}
\par Intelligent reflecting surfaces (IRS), as promising wireless innovations with the development of microelectromechanical systems (MEMS) and metamaterials, have recently been attracting much attention in wireless communications and the Internet of Things (IoT)\cite{MDRenzo, YLiu}. IRSs are two-dimensional, man-made surfaces with low-cost, passive reflecting elements that have adjustable phases and are connected to base stations (BS) or access points (APs) through a smart controller. Specifically, the reflecting elements (e.g., printed dipoles) passively reflect impinging signals without any radio frequency (RF) chains, enabling implementation and management at significantly lower hardware and energy costs compared to traditional active antenna arrays. Additionally, IRSs are characterized by their low profile, lightweight, and conformal geometry, making them easy to mount on or remove from environmental objects for deployment or replacement. Furthermore, IRSs can serve as auxiliary devices in wireless networks and can be easily integrated into them, providing great flexibility and compatibility with existing wireless systems (e.g., cellular or WiFi).

\par In IRS-assisted wireless systems, controlling the phase of reflecting elements is crucial for ensuring optimized performance. Numerous studies have been conducted to determine optimal or suboptimal phases based on various optimization criteria (i.e., spectral efficiency (SE) /energy efficiency (EE) maximization, power minimization, etc.) and various communication scenarios (see, e.g., wireless powered IRS \cite{HXie}, IRS with non-orthogonal multiple access (NOMA) \cite{XLi}, IRS-aided backscatter systems \cite{CZhou}). Besides, there have been also research efforts on investigating the impact of phase errors due to the hardware constraints on the IRS \cite{DLi1, DLi2}, coverage extension via placement optimization \cite{SZeng}, Doppler mitigation for high-speed cases \cite{WWu}, and how to determine the size of the IRS regarding the number configuration with SE and/or EE performance guarantee \cite{DLi3, DLi4}.


\par On the other hand, deep learning (DL) has been widely embraced as a promising solution for addressing the challenge of signaling overhead, as evidenced by numerous studies \cite{CsiNet, XYu, JGuo, DSCsiNet, QCai, XSong}. In \cite{CsiNet}, an autoencoder-based neural network model was proposed to compress channel state information (CSI) in the encoder at the receiver side and reconstruct it in the decoder at the BS side. A novel training strategy was introduced in \cite{JGuo}. Non-local blocks from non-local neural networks were employed in \cite{DSCsiNet} to capture the long-range dependencies of features. In \cite{QCai}, an attention mechanism was integrated into the model to enhance its performance. A self-attention mechanism was adopted in \cite{XSong} to further improve model performance. 
\par The attention mechanism\cite{MHGuo} has evolved into an increasingly significant component of computer vision over the last decade. In existing works\cite{Volodymyr, Vaswani, JHu, SWoo, Vosco, DMisra}, various attention mechanisms have been developed. The squeeze-and-excitation (SE) networks\cite{JHu} have proven to be one of the most popular methods for processing attention with a convolutional operation. SE networks were succeeded by the convolutional block attention module (CBAM)\cite{SWoo}, which focused on providing robust representative attention by incorporating spatial attention and channel attention. CBAM incorporates dimensionality reduction in computing channel attention, which is redundant for capturing nonlinear local dependencies. Unlike traditional SE blocks that 'squeeze' features by utilizing global average pooling (GAP) operations, the tiled squeeze-and-excite (TSE) block employs average pooling to run efficiently on common AI accelerators with data flow design\cite{Vosco}. Additionally, numerical results showed that channel attention learned with local spatial context performs comparably to that learned with global spatial context\cite{Vosco}. By using a triplet attention\cite{DMisra} approach, redundancy was minimized, accounting for cross-dimension interaction efficiently. Triplet attention comprises three branches, each responsible for capturing cross-dimensional relations between spatial and channel dimensions. Convolutional neural networks (CNN) generate multiple output feature maps through convolutions between the same input feature map and multiple kernels\cite{YLeCun}. Given the same input feature map, it is highly likely that each output feature map, along with the channel dimension, is correlated with one another. In other words, the channel dimension information may be related to the spatial dimension information. However, previous attention-based studies calculated attention maps along channels and spatial dimensions separately, with little attention paid to utilizing joint channel and spatial dimension information. To this end, it is interesting to calculate a joint channel and spatial attention map in addition to the conventional attention module, which could enhance the model's performance.

\subsection{Motivation and Novelty}


\par Although the IRS has attracted significant research efforts in the past few years in which quite a number of works investigate on how to compute/optimize the phase shift, most of existing works implicitly assume its availability at the IRS side. Thus, a natural question arises: how to obtain the computed/optimized phase shift for the IRS? This question has received little attention up to now. Moreover, delivering the phase shift information to the IRS is not-trivial due to the potentially huge amount of control signaling overhead (i.e., signaling storm \cite{YChoi, XYi, XLin}). To address this problem, in this paper, we propose two attention-based DL models, namely, the global attention phase shift compression network (GAPSCN) and the simplified GAPSCN (S-GAPSCN).

\par Note that our previous work in \cite{XYu} is the first attempt to solve the problem of phase shift compression, and is highly related with this work. However, we investigate and propose a new method that significantly improves the compression if compared with \cite{ XYu}(but also \cite{CsiNet, JGuo, DSCsiNet, QCai, XSong}). Besides, this method is fundamentally different from existing compression methods in the following aspects:
\begin{itemize}
    \item This work is different from existing state-of-the-art methods for addressing signal overhead challenges in wireless communications (see, e.g. \cite{CsiNet, XYu, JGuo, DSCsiNet, QCai, XSong}) due to the asymmetric architecture design and two novel modules (i.e., global attention mechanism and joint attention-assisted multi-scale network (JAAMSN)).
    \item The proposed global attention mechanism is different from conventional attention mechanisms\cite{JHu,SWoo,Vosco,DMisra} by emphasizing meaningful features along three dimensions: channel dimension, spatial dimension, and joint channel-spatial dimension. 

\end{itemize}

\subsection{Contributions}
The contributions of this paper are summarized as follows:
\begin{itemize}
    \item A novel attention mechanism, namely global attention, was proposed to enhance the model's performance. By investigating the correlating relationship between channel dimension and spatial dimension, global attention is capable of emphasizing a greater number of meaningful features than previous attention works by emphasizing meaningful features along three dimensions: channel dimension, spatial dimension, and joint channel-spatial dimension. 
    \item To address the practical constraints that the IRS controller can not afford high computational complexity, we propose an asymmetric model S-GAPSCN to reduce the computational complexity at the IRS side, in which the architecture of the decoder is significantly simpler than the architecture of the encoder. 
    \item Moreover, to compensate for the performance degradation caused by simplifying the model's architecture, we design a low computation complexity JAAMSN in the decoder of S-GAPSCN. The JAAMSN aims to emphasize meaningful features along the joint channel-spatial dimension and suppress the additive white Gaussian noise (AWGN) effect by adopting a structure of multi-scale and scaling joint attention.
    \item Simulation results demonstrate that the proposed global attention mechanism outperforms the existing attention mechanism by capturing more crucial information. Furthermore, our proposed GAPSCN achieves significant reliability in reconstruction accuracy compared with existing models in wireless communication by emphasizing a great number of meaningful features through the global attention module and alleviating the AWGN effect by the GDN layer. Besides, by using joint attention, JAAMSN achieves a higher performance but at a lower computational cost than the existing attention-guided multi-scale network. In addition, the proposed S-GAPSCN presents an impressive performance at an affordable computational cost due to its asymmetric model structure and JAAMSN.
\end{itemize}

\subsection{Organization and Notation}

\par The remainder of this paper is organized as follows. Section II illustrates the system model. In Section III, we propose an autoencoder-based QPS compression and reconstruction framework. In Section IV, we briefly discuss the global attention mechanism, and the architecture of GAPSCN and compare the performance among the proposed global attention mechanism and existing works. An asymmetric model, S-GAPSCN, is presented in Section V. In Section VI, we provide details regarding the training procedure, simulation results, and an analysis of the simulation results. In Section VI, we conclude this paper. 
\par The notations used in our paper are listed as follows. The superscript $H$ is used to represent the conjugate transpose. Term $\mathbb{C}$ denotes the complex number. $\mathcal{CN}(\mu,\sum)$ is the circularly symmetric complex Gaussian (CSCG) distribution where $\mu$ and $\sum$ are the mean vector and the covariance matrix, respectively. $\lVert \cdot \rVert ^2$ is the norm of an input vector.

\section{System Model}

\begin{figure}[t]
\centering
\includegraphics[scale=0.5]{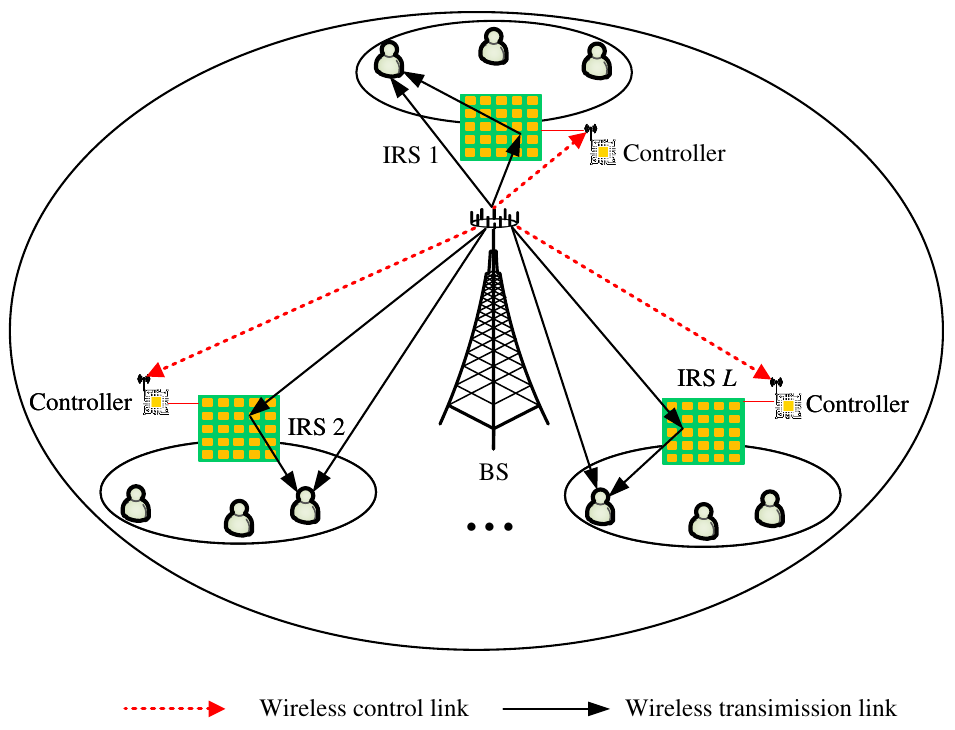}
\caption{An IRS-assisted wireless system.}
\label{IRS}
\end{figure}

\par In this paper, as shown in Fig.~\ref{IRS}(a), we consider the IRS-assisted downlink system in a single-cell network where $\textit{L}$ IRSs are deployed to assist in the communications from multi-antenna BS to $\textit{K}$ multi-antenna users. Specifically, the BS, each user, and each IRS are equipped with an $N_t$-element antenna array, an $N_r$-element antenna array, and an $M$-element antenna array respectively.
The baseband equivalent channels from the AP to $\textit{l}$-th IRS, from
the $\textit{l}$-th IRS to $\textit{k}$-th user, and from the BS to $\textit{k}$-th user are denoted by
$\textbf{G}_{l}\in \mathbb{C}^{M \times N_t}$, $\textbf{h}_{r,l,k}\in \mathbb{C}^{M \times N_r}$, and $\textbf{h}_{d,l,k}\in \mathbb{C}^{N_r \times N_t}$, respectively, where $l = 1, \cdots, L$ and $k = 1, \cdots, K$. By denoting $P$ and $s$ as the transmit power and the transmitted signal of $S$, the received signal reflected by the $l$-th IRS received at $k$-th user from both the BS-user and BS-IRS-user channels is then expressed as
\begin{equation}
    \label{eq1}
    y_d = \sqrt{P}\textbf{h}^H_{r,l,k}\Phi \textbf{G}_{l}s+\sqrt{P}\textbf{h}_{d,l,k}s+u_d,
\end{equation}
where $\Phi= \rho \rm{diag}(e^{j \theta_1},e^{j \theta_2},\ldots,e^{j \theta_M}) \in \mathbb{C}^{M \times M}$ is the phase matrix ($j$ is the imaginary unit) with $\rho \in (0,1]$ and $\left\{ \theta_m \right\}^M_{m=1}$ is the reflection coefficient and phases of the IRS, and $u_d \sim \mathcal{CN}(0,\sigma_d^2)$ is the additive white Gaussian noise (AWGN). However, due to the finite resolution of the IRS, $\theta_m$ can only take a finite number of discrete values (see, e.g., \cite{DLi1, DLi2}), i.e., $2^K$ quantization levels and $K$ denotes the number of quantization bits. Therefore, the phase $\theta_m$ can be uniformly quantized by $K$ bits by using, then the set of quantized phase shift (QPS) is given by $\left\{ 0, \frac{2\pi}{2^K}, \cdots,\frac{(2^K-1)2\pi}{2^K} \right\}$. 

\par $\textit{Remark 1}$: It should be noted that in the IRS-assisted wireless system, the channels for data transmission and control signaling transmission are different\cite{XLin}. For example, in 5G new radio (NR), the channel for downlink data transmission is the physical downlink shared channel (PDSCH), while the channel for downlink control signaling transmission is the physical downlink control channel (PDCCH). With these two separate transmission channels, we can focus on addressing the QPS signaling overhead without worrying about data overloading. Moreover, the size of the QPS is only related to the number of reflecting elements and the quantization level.

\begin{figure*}[t]
\centering
\includegraphics[scale=0.65]{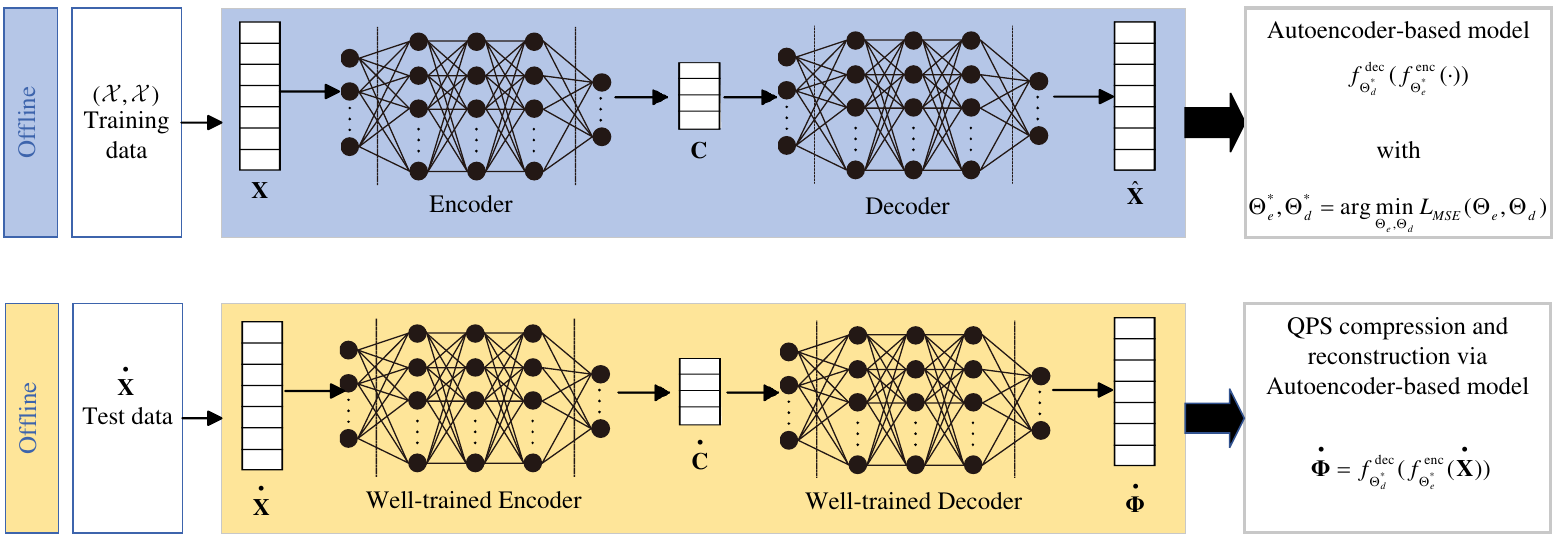}
\caption{An autoencoder-based QPS compression and reconstruction framework.}
\label{framework}
\end{figure*}

\section{Autoencoder-based QPS Compression and Reconstruction Framework}

\par In this paper, we consider the QPS signaling compression from the receiver to the IRS to deliver the QPS by designing an autoencoder-based framework, which consists of an offline training phase and an online compression and reconstruction phase, as illustrated in Fig.~\ref{IRS}(b). To be specific, the autoencoder is composed of two components: an encoder and a decoder\cite{Goodfellow}. The encoder is equipped at BS aiming at compressing the original QPS to a lower dimension. And the decoder which is located at the IRS side reconstructs the original QPS. For the online training phase, a DL model training process is operated to obtain a well-trained autoencoder-based model. For the online QPS compression and reconstruction phase, the QPS for the test is sent to the encoder of the well-trained autoencoder-based model to obtain the compressed QPS. Then, the compressed QPS will be sent to the decoder of the well-trained autoencoder-based model through a control channel to reconstruct the QPS. The details of offline training and online compression and reconstruction will be elaborated in the following.

\subsection{Offline Training}

\par For giving a training data set 
\begin{equation}
    (\mathcal{X}, \mathcal{X}) = \{(\mathbf{x}^{(1)},\mathbf{x}^{(1)}),\cdots , (\mathbf{x}^{(t)},\mathbf{x}^{(t)})\} ,
\end{equation}
where $(\mathbf{x}^{(t)},\mathbf{x}^{(t)})$, $n\in \{1,\cdots,T\}$ are the input data and the ground truth of the $t$-th of $(\mathcal{X}, \mathcal{X})$. The input data and the ground truth are the same since the task aims at receiving the original QPS based on the compressed QPS. 

\par Note that the encoder and the decoder of the autoencoder-based model can be built by various layers and mechanisms. e.g., fully connected layer, convolutional neural network (CNN), recurrent neural network (RNN), attention mechanism, etc. Denote by $f^{\text{enc}}_{\mathbf{\Theta_e}}(\cdot)$ and $f^{\text{dec}}_{\mathbf{\Theta_d}}(\cdot)$ the expression of the encoder and the decoder, respectively, where $\Theta_e$ and $\Theta_d$ are the parameters of the encoder and the decoder, respectively. Then the output of the autoencoder-based model can be expressed as:
\begin{equation}
    \hat{\textbf{X}} =  f^{\text{dec}}_{\mathbf{\Theta_d}}(\textbf{C}) = f^{\text{enc}}_{\mathbf{\Theta_e}}(\textbf{X}),
\end{equation}
where $\textbf{C}$ is the compressed input data and $\hat{\textbf{X}}$ is reconstructed input data. We choose the mean square error (MSE) as our cost function according to the minimum mean square error (MMSE) criterion which can be expressed as
\begin{equation}
 \begin{split}
  L_{MSE}(\Theta_e,\Theta_d) &= \frac{1}{T}\sum^T_{t=1}  \lVert \mathbf{X}^{(t)} - \hat{\textbf{X}}^{(t)} \rVert ^2 \\
    &= \frac{1}{T}\sum^T_{t=1}  \lVert \mathbf{X}^{(t)} - f^{\text{dec}}_{\mathbf{\Theta_d}}(f^{\text{enc}}_{\mathbf{\Theta_e}}(\textbf{X})) \rVert ^2.    
\end{split}   
\end{equation}
The autoencoder-based model utilizes the backpropagation (BP) algorithm to progressively update the network parameters to finally obtain the well-trained model. Then, the well-trained autoencoder-based model can be expressed as
\begin{equation}
    \hat{\textbf{Z}} = f^{\text{dec}}_{{\Theta_d^*}}(f^{\text{enc}}_{{\Theta_e^*}}(\textbf{Z})),
\end{equation}
where $\textbf{Z}$ is an arbitrary input, $\Theta_d^*$ and $\Theta_e^*$ denote the well-trained parameters by minimizing $L_{MSE}(\Theta_e,\Theta_d)$.

\subsection{Online Compression and Reconstruction}

\par As shown in Fig.~\ref{framework}, the given test data set $\Dot{\mathbf{X}}$ is sent to the well-trained autoencoder-based model as an input to process the QPS compression and reconstruction. Then the reconstructed QPS $\Dot{\mathbf{\Phi}}$ based on the well-trained autoencoder-based model can be expressed as
\begin{equation}
    \Dot{\mathbf{\Phi}} = f^{\text{dec}}_{{\Theta_d^*}}(f^{\text{enc}}_{{\Theta_e^*}}(\Dot{\mathbf{X}})),
\end{equation}

\subsection{Algorithm for QPS Compression and Reconstruction}

\par Based on the above analysis, we then summarize the proposed autoencoder-based QPS compression and reconstruction framework as an algorithm as shown in Algorithm 1, where we use $i$ and $I$ to denote the iteration index and the maximum iteration number, respectively.

\par \textit{Remark 2}: It should be noted that this work can be easily extended to any control signaling compression task without requiring changes to the protocol, CSI signaling overhead, or QPS signaling overhead.

\begin{algorithm}[t]
\caption{Algorithm for QPS Compression and Reconstruction}
\label{alg:cap}
\begin{algorithmic}
\State \textbf{Initialization}:$i$=0
\State \textbf{Offline Training phase}:
\State 1:\quad \quad \textbf{Input}: Training set $(\mathcal{X}, \mathcal{X})$
\State 2:\quad \quad ~ ~\textbf{while}~$i\leq I$~\textbf{do}
\State 3:\quad \quad ~ ~ Update $\Theta_e$ and $\Theta_d$ by BP algorithm to minimize 
\State \quad \quad \quad \quad \quad$L_{MSE}(\Theta_e,\Theta_d)$
\State \quad\quad \quad \quad~$i=i+1$
\State 4:\quad \quad ~ ~\textbf{end while}
\State 5:\quad \quad \textbf{Output}: A Well trained autoencoder-based model 
\State \quad \quad \quad$f^{\text{dec}}_{{\Theta_d^*}} (f^{\text{enc}}_{{\Theta_e^*}}(\mathbf{\cdot}))$ with optimized weight $\Theta_d^*$ and $\Theta_e^*$.
\State \textbf{Online Compression and Reconstruction phase}:
\State 6:\quad \quad \textbf{Input}: Testing set $\Dot{\mathbf{X}}$
\State 7:\quad \quad ~ ~\textbf{do}~$f^{\text{dec}}_{\mathbf{\Theta_d^*}}(f^{\text{enc}}_{\mathbf{\Theta_e^*}}(\Dot{\mathbf{X}}))$
\State 8:\quad \quad \textbf{Output}: Reconstructed QPS $\Dot{\mathbf{\Phi}}$.
\end{algorithmic}
\end{algorithm}

\section{GAPSCN for QPS Compression and Reconstruction} 

\par In this paper, we propose a novel global attention mechanism that enables us to emphasize a greater number of meaningful features than conventional attention mechanisms. In this section, we begin with an introduction to the preliminary knowledge of conventional neural network layers and activation functions. Then, we introduce the existing attention mechanisms and elaborate on the proposed global attention mechanism. Moreover, we briefly describe the architecture of the proposed GAPSCN. Finally, we provide a performance comparison between the proposed global attention mechanism and existing attention mechanisms, along with three ablation studies, by evaluating the mean square function and the accuracy function based on TensorFlow\cite{TensorFlow}.

\subsection{Preliminary Knowledge}

\subsubsection{Convolutional Neural Network}
\par A CNN \cite{YLeCun} is a feed-forward neural network that is generally used to analyze visual images by processing data with a grid-like topology, which has achieved tremendous success in numerous fields, such as image classification and image segmentation. The CNN employs a mathematical operation called convolution, which is a specialized kind of linear operation. Given an input $X$ and a kernel $K$, the i-th row and the j-th column output of convolution can be expressed as\cite{Goodfellow}:
\begin{equation}
y(i,j) = (K\ast X) = \sum_m \sum_n x(i-m,j-n)k(m,n).
\end{equation}

\subsubsection{Activation function}
Since the task is often a nonlinear mapping task, we expect the mapping from the CNN input to its output to also be highly nonlinear. To this end, we increase the nonlinearity of the CNN by activating the CNN with a rectified linear unit (ReLU), which can be expressed as:
\begin{equation}
\sigma_r(x) = max \{0,x \}.
\end{equation}
There is no parameter inside a ReLU layer, and hence no need for parameter learning in this layer.
The sigmoid activation function is expressed as:
\begin{equation}
\sigma_s(x) = \frac{1}{1+e^{-x}},
\end{equation}
where $\sigma_s(x) \in (0,1)$. Activated by the sigmoid function, the output will be nonlinearly transformed to the range of $(0,1)$.

\subsubsection{Generalized divisive normalization layer} 
\par GDN is defined in terms of an inevitable nonlinear transformation that is optimized so as to Gaussianize the data. The transformation process is given by
\begin{align}
  &y_i=\frac{z_i}{(\beta_i+\sum_j \gamma_{ij}|z_j|^ {\alpha_{ij}}) ^ {\epsilon_i}},\\
 &z_i=Hx_i,  
\end{align}
where $x$ is the input vector, $\beta, \epsilon$ are vector parameters, and $\alpha,\gamma$ are matrix parameters. The GDN transformation can be efficiently inverted using a fixed point iteration. The GDN/IGDN is an efficient algorithm for fitting the parameters of this transformation, minimizing the Kullback-Leibler divergence of the distribution of transformed data against a Gaussian target, which preserves better original data distribution information than ReLU.

\begin{figure*}[t]
\centering
\includegraphics[width=7in]{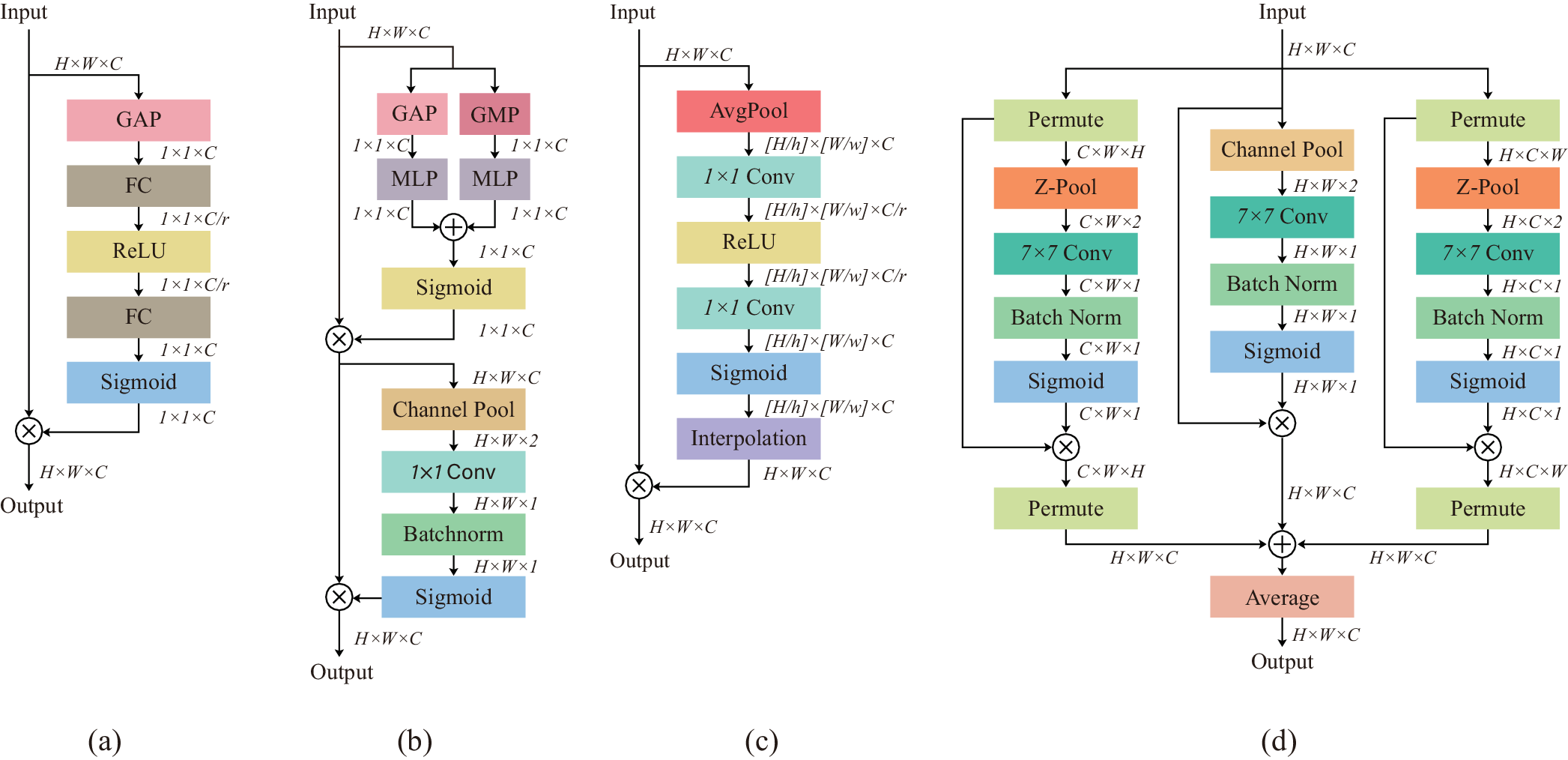}
\caption{Comparison with different attention mechanisms: (a) SE block; (b) CBAM; (c) TSE block; (d) Triplet attention; $1\times1$ Conv and $7\times7$ Conv denote the convolutional layer with kernel size $1\times1$ and $7\times7$, respectively. The operation of channel pool and z-pool process a GAP and a GMP to the last dimension of the input, respectively, and concat the output along the last dimension.}
\label{attention}
\end{figure*}

\subsection{Existing Attention mechanisms}

\par As a first step in our explanation of global attention, let us review existing attention mechanisms, i.e., SE block\cite{JHu}, CBAM\cite{SWoo}, TSE block\cite{Vosco} and triplet attention\cite{DMisra}, which are illustrated in Fig.~\ref{attention}(a), Fig.~\ref{attention}(b), Fig.~\ref{attention}(c) and Fig.~\ref{attention}(d), respectively. 

\par \subsubsection{SE block} The SE block compute the attention map $AM$ by utilizing a GAP to capture the channel-wise information and a fully connected (FC) layer to reduce the computational complexity followed by a nonlinear activation function ReLU. Then, it uses a FC layer to reconstruct the original dimension, and a sigmoid function used for nonlinear transformation of the information to interval $[0,1]$. At last, the attention process computes the element-wise product of the input and the attention map. The attention is computed as:
\begin{equation}
    \begin{split}
&AM_{se} = \sigma(f_{fc}(f_{ReLU}(f_{fc}((f_{GAP}(X))))),\\
&ATT_{se} = X \otimes AM_{se},
    \end{split}
\end{equation}
where $X$, $\sigma$, $\otimes$, $f_{fc}$, $f_{Relu}$ and $f_{GAP}$ denote the input, the sigmoid function, element-wise product, fully connect layer, nonlinear activation function ReLU and GAP operation, respectively.

\par \subsubsection{CBAM} The CBAM consists of two modules, one for channel attention and the other for spatial attention. By using GAP and global max pooling (GMP), channel attention produces two spatial features. Then, two features are forwarded to a multilayer perceptron (MLP) with one hidden layer, and the output features are then merged using the element-wise summation. Let $AM_c$ be the channel attention map produced by applying a sigmoid function to the summation feature. The channel attention is computed as:
\begin{equation}
    \begin{split}
     &AM_c = \sigma(f_{MLP}(f_{GAP}(X))+f_{MLP}(f_{GMP}(X))),\\
    &ATT_c = X \otimes AM_c ,       
    \end{split}
\end{equation}
where  $f_{MLP}$ and $f_{GMP}$ denote the MLP layer and GMP operation, respectively. 
\par Different from channel attention, the two spatial features are concatenated together and then forwarded to a convolutional layer to produce the spatial attention map $AM_s$. The spatial attention is computed as:
\begin{equation}
    \begin{split}
&AM_s = \sigma(f_{BN}((f_{1\times1}([f_{GAP}(Att_c);f_{GMP}(Att_c)])))),\\
&ATT_s = ATT_c \otimes AM_s,    
    \end{split}
\end{equation}
where $f_{BN}$ and $f_{1\times1}$ denote the batch normalization layer and convolutional layer with the kernel size $1\times1$, respectively.

\par \subsubsection{TSE block} The only difference between traditional SE block and TSE block is the structure of the 'squeeze'. Different from traditional SE block using a GAP, the TSE block utilizes an average pooling to run efficiently on common AI accelerators with data flow. The attention map is calculated in the TSE block by an average pooling layer, a convolutional layer with a nonlinear activation function ReLU, a convolutional layer with a nonlinear activation function sigmoid and a nearest-neighbor interpolation. The attention is computed as:
\begin{equation}
    \begin{split}
&AM_{tse} = f_{nnp}(\sigma(f_{1\times1}(f_{ReLU}(f_{1\times1}(f_{ap}(X)))))),\\
&ATT_{tse} = X\otimes AM_{tse},
    \end{split}
\end{equation}
where $f_{nnp}$ and $f_{ap}$ denote the nearest-neighbor interpolation and the average pooling operation, respectively.

\begin{figure*}[t]
\centering
\includegraphics[scale=0.7]{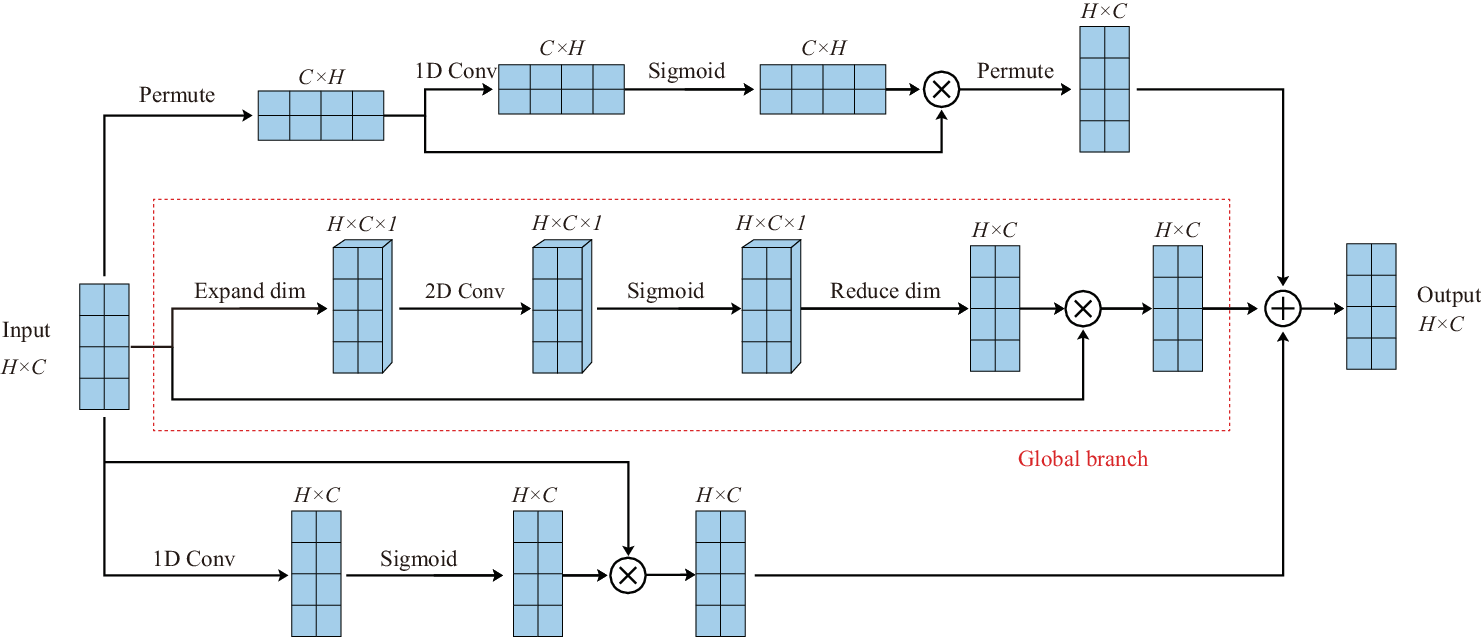}
\caption{The architectures of the global attention. Expand dim and Reduce dim denote the operation of expanded dimension and the operation of reduced dimension, respectively.}
\label{global}
\end{figure*}

\par \subsubsection{Triplet Attention} This approach consists of three branches: the first and the third branches are similar to spatial attention in CBAM except a rotation operation added to the beginning of both branches; while the second branch is a direct equivalent to the spatial attention in CBAM. The final step involves applying a simple average operation to the output of three branches. The triplet attention can be computed as:
\begin{equation}
    \begin{split}
    &AM_{ta1} = \sigma(f_{7\times7}([f_{GAP}(f_{Rt}(X));f_{GMP}(f_{Rt}(X))])),\\
    &ATT_{ta1} = f_{Rt}(f_{Rt}(X) \otimes AM_{ta1}),   \\
    &AM_{ta2} = \sigma(f_{7\times7}([f_{GAP}(f_{Rt}(X));f_{GMP}(f_{Rt}(X))])),\\
    &ATT_{ta2}= f_{Rt}(f_{Rt}(X) \otimes AM_{ta2}),   \\
    &AM_{ta3} = \sigma(f_{7\times7}([f_{GAP}(F);f_{GMP}(X)])),\\
    &ATT_{ta3} = X \otimes AM_{ta3},   \\
    &ATT_{ta} = \frac{1}{3}(Att_{ta1}+Att_{ta2}+Att_{ta3}),        
    \end{split}
\end{equation}
where $f_{1\times1}$ and $f_{Rt}$ denote the convolutional layer with the kernel size $7\times7$ and the operation of permutation, respectively.

\subsection{The Proposed Global Attention Mechanism}

\par The output feature maps in a CNN layer are generated by a convolution between the input feature map and multiple kernels. With the same input feature map, each of the output feature maps along the channel dimension in the CNN is correlated with each other. In other words, for the output feature maps of the CNN layer, there may exist a correlation between the channel dimension information (the last dimension of the feature maps) and the spatial dimension information (the dimensions excluding the last dimension of the feature maps).

\par Most of existing attention-based works, crucial information has been emphasized along two dimensions separately: channel dimension (what meaningful information requires attention) and spatial dimension (where meaningful information requires attention). However, the channel dimension and spatial dimension are generally considered separately in previous works when processing the attention map, where the correlation between the channel dimension and the spatial dimension is neglected, and thus the joint information emphasizing is missing.

\par Based on this observation, we propose a novel global attention mechanism by calculating attention maps not only across the channel dimension or the spatial dimension separately but also the joint channel dimension and spatial dimension globally.
Besides, from the empirical results in \cite{Vosco}, we abandon the use of the pooling operation. In the following section, we will examine the ablation study of the information replenishment. 

\begin{figure*}[t]
\centering
\includegraphics[scale = 0.9]{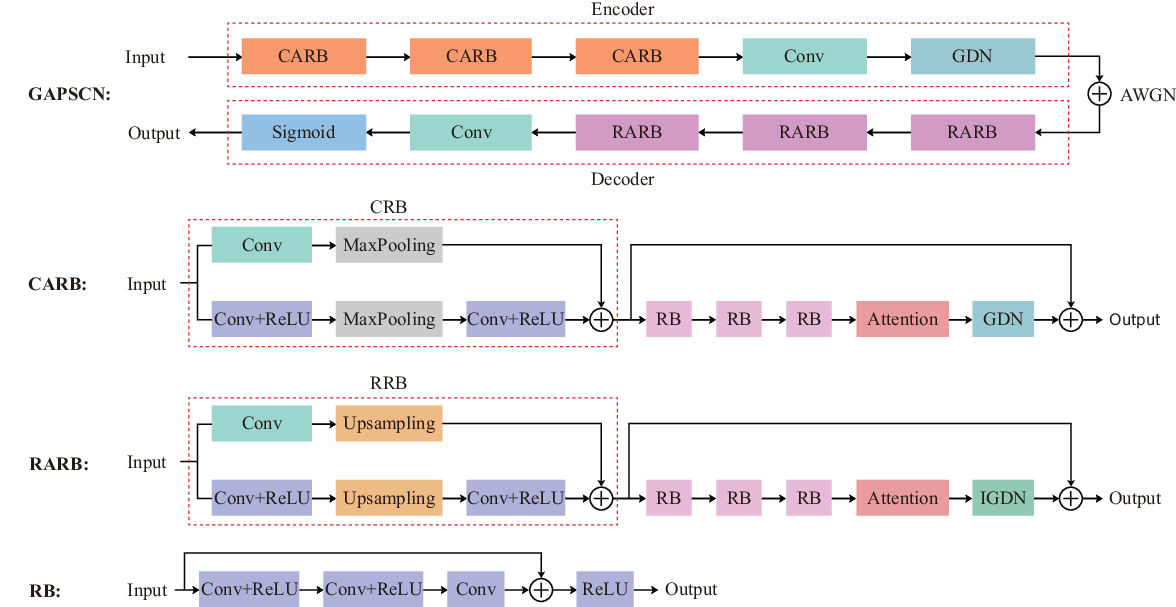}
\caption{The architectures of the GAPSCN.}
\label{GAPSCN}
\end{figure*}

\par The architecture of the global attention mechanism, as illustrated in Fig.~\ref{global}, comprises three distinct branches. The lower branch calculates the attention map along the spatial dimension by utilizing a one-dimensional (1D) convolutional layer with a kernel size of 1. Subsequently, the output feature map undergoes activation through a non-linear sigmoid function. This sigmoid function enables a non-linear transformation, constraining the feature map values within the $(0,1)$ range. Furthermore, the element-wise product (also referred to as the Hadamard product) is generated between the input and feature map, assigning weights close to 1 for essential features and approaching 0 for less significant features. The upper branch computes attention along the channel dimension by incorporating a permutation layer, a 1D convolutional layer with a kernel size of 1, and a non-linear sigmoid function. After performing the element-wise product between the attention map input and the rotated dimension input, the output dimension is rotated back to the original input dimension.
In the middle branch, to holistically integrate the channel and spatial dimensions, an additional null dimension is added to the input (e.g., the input dimension is $H\times C$, while the output following the expansion operation is $H\times C\times1$). A two-dimensional (2D) convolutional layer with a kernel size of $1\times1$ and a non-linear sigmoid activation function are subsequently employed to compute the attention map. To maintain consistent dimensions of the attention map, the squeeze dimension operation is applied before executing the element-wise product between the attention map and the expanded dimension input. As a result, the global attention mechanism captures dependencies between the ($C$, $H$), ($H$, $C$), and ($H\times C$, $H \times C$) dimensions of the input, respectively. Finally, to achieve a comprehensive attention map, the outputs from each branch are aggregated. Therefore, the global attention mechanism can be expressed as
\begin{equation}
    \begin{split}
&AM_{ga1} = \sigma(f_{1}(f_{Rt}(X)),\\
&ATT_{ga1} = f_{Rt}(X) \otimes AM_{ga1}, \\
&AM_{ga2} = \sigma(f_{rd}(f_{1\times1}(f_{ep}(X)))),\\
&ATT_{ga2} = f_{ep}(X) \otimes AM_{ga2},   \\    
&AM_{ga3} = \sigma(f_{1}(X)),\\
&ATT_{ga3} = X \otimes AM_{ga3},  \\
&ATT_{ga} = ATT_{ga1}+ATT_{ga2}+ATT_{ga3},    
    \end{split}
\end{equation}
where $f_1$, $f_{ep}$ and $f_{rd}$ denote the convolutional layer with the kernel size 1, the operation of the expanded dimension and the operation of the reduced dimension, respectively.

\subsection{The Architecture of Proposed GAPSCN}

\par The architecture of GAPSCN is shown in Fig.~\ref{GAPSCN}, which is composed of an encoder and a decoder. The encoder is comprised of multiple compression attention residual blocks (CARBs), one convolutional layer, and one GDN layer\cite{BJohannes}, while the decoder is comprised of multiple reconstruction attention residual blocks (RARBs) and one convolutional layer. (The number of CARBs and RARBs depends on the compression rate (CR). For instance, GAPSCN with $CR = \frac{1}{8}$ has three CARBs and three RARBs.

\par \par The CARB is composed of the compression residual blocks (CRBs), followed by the residual blocks (RBs), a global attention module, and a GDN layer. As shown in Fig.~\ref{GAPSCN}, the CRB is made up of two branches: the first branch is composed of a convolutional layer followed by a downsampling layer (max pooling). The second branch consists of a convolutional layer followed by a downsampling layer and a convolutional layer. The output of each branch is then combined. The last step involves transforming the input distribution into a Gaussian distribution using a GDN layer. Based on the success in \cite{ZCheng}, we apply three RBs to capture deep information before the attention module. Every RB consists of a 1D convolutional layer with the kernel size 1, followed by a 1D convolutional layer with the kernel size 3, and finally a 1D convolutional layer with the kernel size 1. At the end of the RB, a residual add operation is applied between the input and output of the last convolutional layer. By utilizing the GDN layer, the output of the attention module will be converted to a Gaussian distribution. Last, we adopt the skip connection, which adds the output of the RBs and the output of the GDN layer for faster convergence\cite{KHe}.

\par The RARB is composed of a reconstruction residual block (RRB), three RBs, a global attention module, and an IGDN layer. As shown in Fig.~\ref{GAPSCN}, the RRB is almost equivalent to the CRB, the main difference is that the downsampling layer is switched for an upsampling layer (double the input along the first dimension) and the IGDN layer is used to transform the Gaussian distribution back to the original uniform distribution. To capture the deep information of reconstructed features, we use three RBs similar to CARB. In the following step, we convert the output of the attention module to the original input distribution by utilizing the IGDN layer. Last but not least, in RARB, the RBs and the IGDN layer outputs will be combined for faster convergence\cite{KHe}.

\par The CARB's output will go through a convolutional layer with the kernel size 1 and a filter size 1 to reduce the dimension of the channel to 1. Thereafter, a GDN layer is applied to transform the uniform data distribution to Gaussian distribution. The output of a RARB will go through a convolutional layer with the kernel size 1 and a filter size 1 and is activated by a nonlinear function sigmoid in the decoder in order to reconstruct the QPS. A comprehensive view of GAPSCN can be obtained by noting the input/output shape of each layer as shown in Table~\ref{shape}.

\begin{table}[t]
\caption{Size of each input/output of each layer/block}
\centering
\begin{tabular}{|c|c|c|}
\hline
\multicolumn{3}{|c|}{\textbf{Input}: QPS with size $MK\times 1$ and $O$ neurons} \\
\hline
Layer/Block &Input size & Output size  \\
\hline
\multicolumn{3}{|c|}{\textbf{Encoder}} \\
\hline
CARB & $MK \times 1$ & $(MK/2)\times O$\\
  \hline
CARB & $(MK/2) \times O$ & $(MK/4) \times O$\\
  \hline
CARB & $ (MK/4) \times O $ & $ (MK/8) \times O $\\
  \hline
GDN & $ (MK/8) \times O $ & $ (MK/8) \times O $\\
  \hline
Conv & $ (MK/8) \times O $ & $ (MK/8) \times 1 $\\
  \hline 
GDN & $ (MK/8) \times 1 $ & $ (MK/8) \times 1 $\\
  \hline 
\multicolumn{3}{|c|}{\textbf{Decoder}} \\
\hline
RARB & $ (MK/8) \times 1 $ & $ (MK/4) \times O $\\
  \hline
RARB & $ (MK/4) \times O $ & $ (MK/2) \times O $\\
  \hline
RARB & $ (MK/2) \times O $ & $ MK \times O $\\
  \hline
IGDN & $ MK \times O $ & $ MK \times O $\\
  \hline
Conv & $ MK \times O $ & $ MK \times 1 $\\
  \hline
Sigmoid & $ MK \times 1 $ & $ MK \times 1 $\\
  \hline
\multicolumn{3}{|c|}{\textbf{Output}: Reconstructed QPS with size $MK\times 1$} \\
\hline
\end{tabular}
\label{shape}
\end{table}
\begin{figure*}[t]
\centering
\subfloat[]{\includegraphics[width=2.5in]{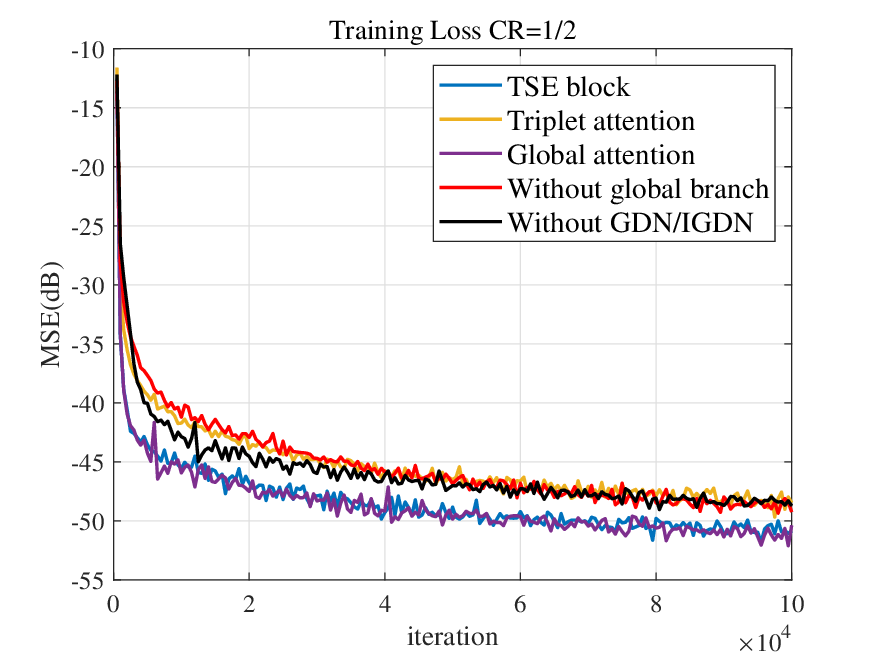}%
\label{tacc12}}
\hfil
\hspace{-8.6mm}
\subfloat[]{\includegraphics[width=2.5in]{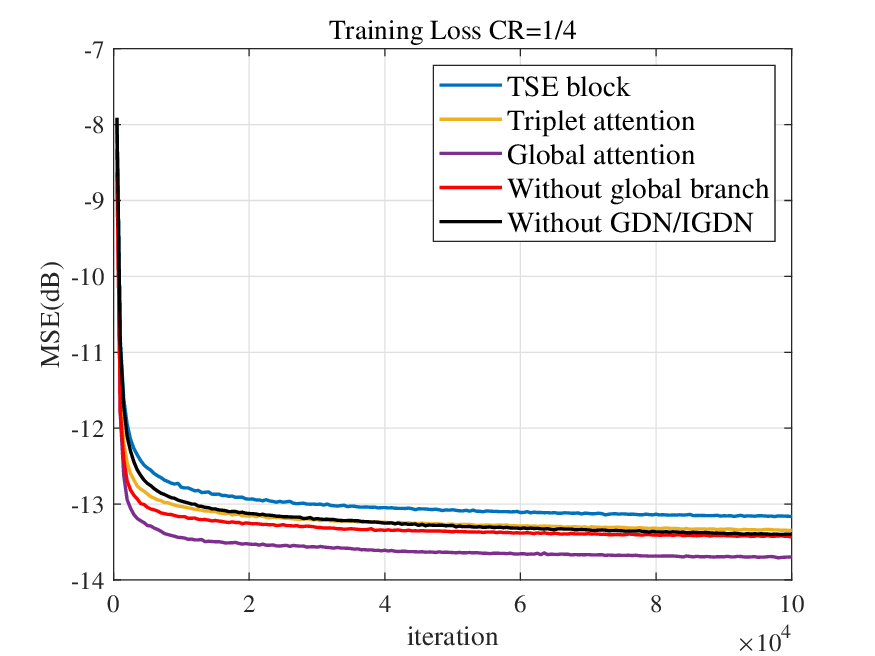}%
\label{tacc14}}
\hfil
\hspace{-8.6mm}
\subfloat[]{\includegraphics[width=2.5in]{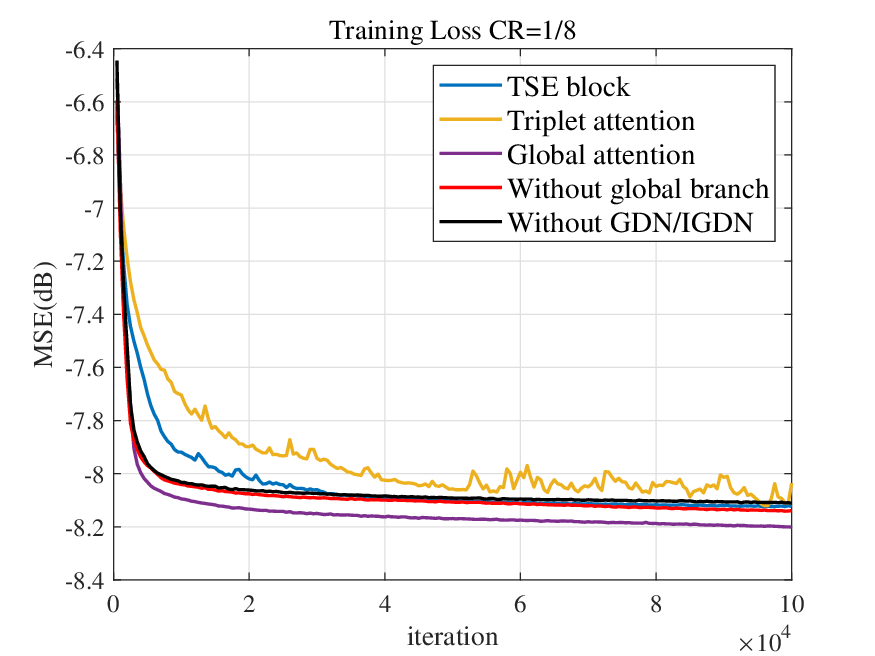}%
\label{tacc18}}
\hfil
\hspace{-8.6mm}
\subfloat[]{\includegraphics[width=2.5in]{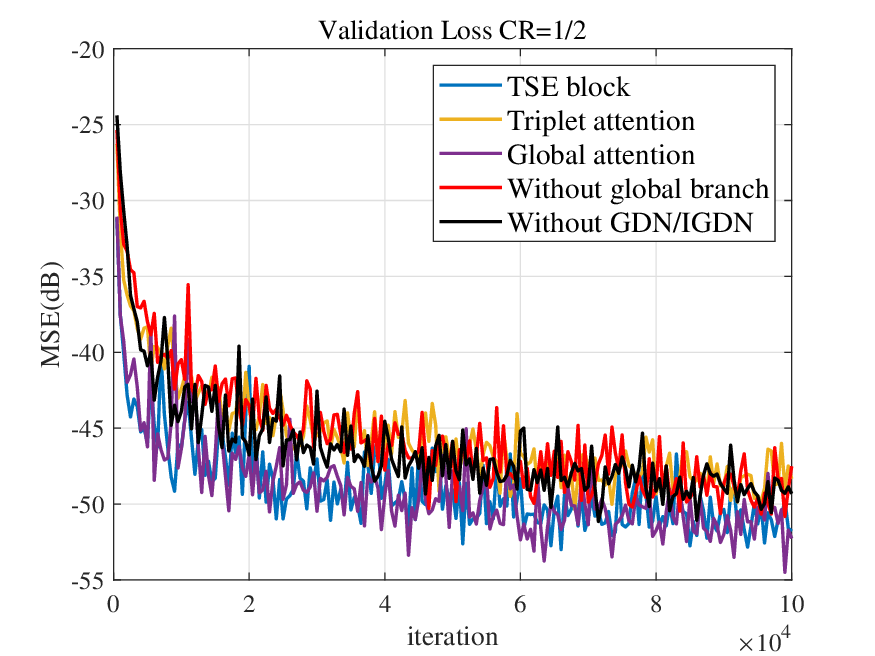}%
\label{vacc12}}
\hfil
\hspace{-8.6mm}
\subfloat[]{\includegraphics[width=2.5in]{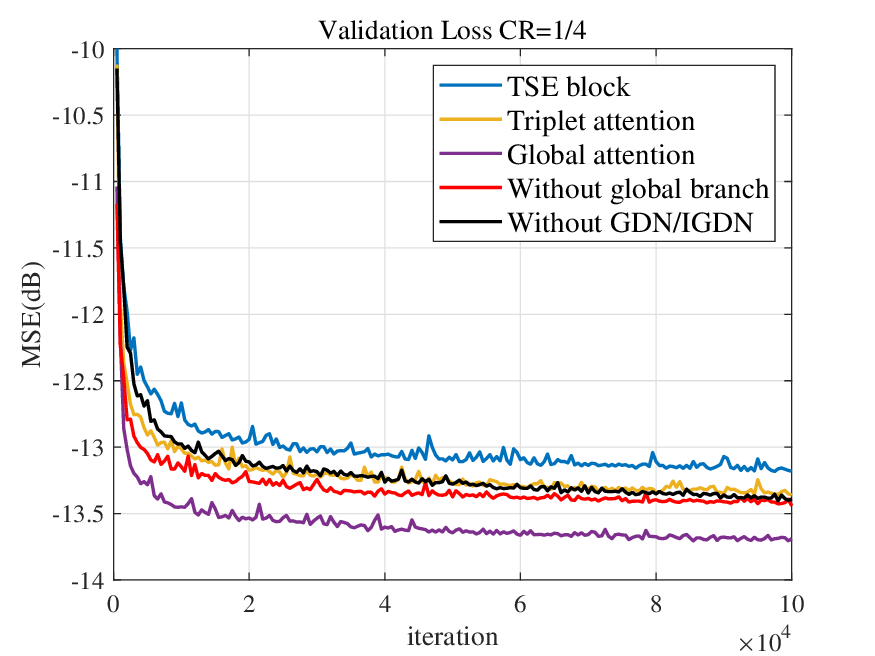}%
\label{vacc14}}
\hfil
\hspace{-8.6mm}
\subfloat[]{\includegraphics[width=2.5in]{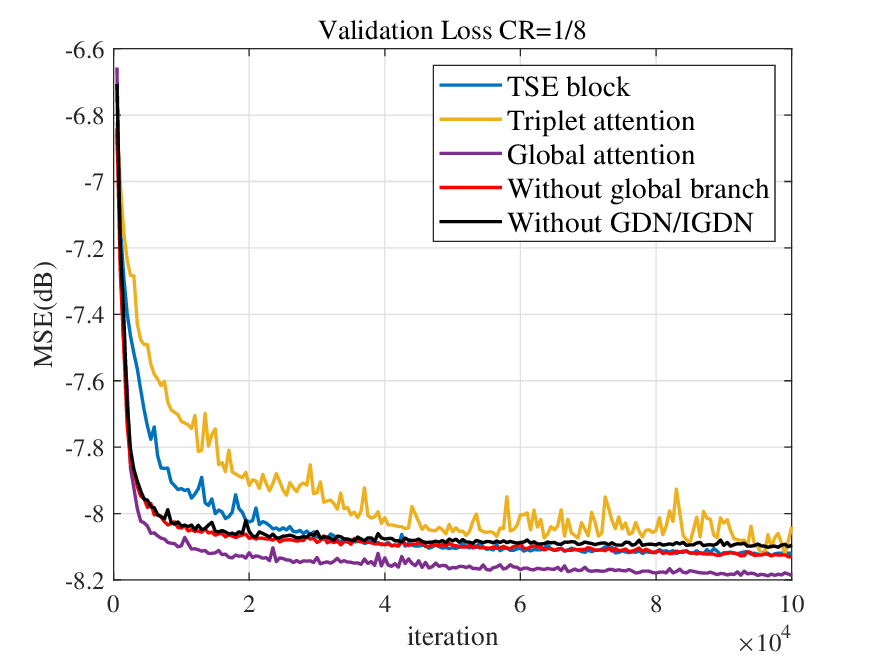}%
\label{vacc18}}
\caption{Training Loss and Validation Loss of GAPSCN + various attention mechanisms (global attention, global attention without global branch, TSE block and triplet attention) and GAPSCN without GND/IGDN layer under different CR. (a) Training accuracy under $CR=\frac{1}{2}$. (b) Training accuracy under $CR=\frac{1}{4}$. (c) Training accuracy under $CR=\frac{1}{8}$. (d) Validation accuracy under $CR=\frac{1}{2}$. (e) Validation accuracy under $CR=\frac{1}{4}$. (f) Validation accuracy under $CR=\frac{1}{8}$.}
\label{fig_sim}
\end{figure*}

\begin{table*}[ht]
\caption{Training accuracy and Validation accuracy}
\centering
\begin{tabular}{|c|c|c|c|c|c|c|c|c|c|}
\hline
\multirow{2}{*}{Type of Module}&\multicolumn{2}{|c|}{\textbf{$CR=\frac{1}{2}$}} &\multicolumn{2}{|c|}{\textbf{$CR=\frac{1}{4}$}} &\multicolumn{2}{|c|}{\textbf{$CR=\frac{1}{8}$}}\\
\cline{2-7}
 &  Training acc  & Validation acc  &  Training acc & Validation acc  &  Training acc & Validation acc  \\
\hline
TSE block\cite{JHu} & 0.999992& 0.999994 & 0.9307 & 0.9309 & 0.756 & 0.756 \\
\hline
Triplet attention\cite{DMisra} & 0.999986 & 0.999988 & 0.9353 & 0.9354 & 0.752 & 0.752 \\
\hline
Global attention & 0.999992 & 0.999995 & 0.9409 & 0.9410 & 0.768 & 0.768 \\
\hline
Without global branch & 0.999986 & 0.999990 & 0.9368 & 0.9369 & 0.765 & 0.764 \\
\hline
Without GDN/IGDN & 0.999985 & 0.999991 & 0.9354 & 0.9353 & 0.763 & 0.762 \\
\hline
\end{tabular}
\label{par}
\end{table*}

\subsection{Performance Comparison and Ablation Study}

\par To determine the relative effectiveness of different attention modules, we chose the GAPSCN as the basic architecture and train it by switching between different attention modules. In addition, we conduct two ablation studies here to study the effect of the GDN/IGDN layer, global branch. We only chose the triplet attention and the TSE block as the comparison candidates since triplet attention already outperforms CBAM, and  TSE block outperforms SE block. The CBAM combined with attention CsiNet\cite{QCai} and the SE block combined with SALDR\cite{XSong} will be compared with the proposed GAPSCN in Section VI. Due to the 1D nature of input data, the 2D convolutional layer in each attention module will be converted to a 1D convolutional layer. We will elaborate the hyperparameter settings as well as the training procedure in Section VI.

\par As shown in Fig.~\ref{fig_sim}, the training and validation loss for different attention mechanisms are illustrated (global attention, global attention without the global branch, TSE block and triplet attention) as well as GAPSCN without GND/IGDN layer under the conditions of CR$\in[\frac{1}{2},\frac{1}{4},\frac{1}{8}]$. With an increasing CR, the training loss and validation loss of each model are decreasing. Across all CRs, it is found that global attention is superior both in terms of training loss and validation loss, demonstrating the effectiveness of correlated information replenishment. In addition, by alleviating the AWGN effect, compression performance can be improved. It has been demonstrated that GAPSCN with GND/IGDN layers achieves a much lower training loss and validation loss than GAPSCN without GND/IGDN layers.

\par To further evaluate the performance, we select the accuracy function provided by TensorFlow. The training accuracy and validation accuracy of each model are presented in Table~\ref{par}. The global attention achieves the highest training accuracy and corresponding validation accuracy in each CR, with a similar fashion to the trends in training loss and validation loss. Both ablation studies show trends that are similar to MSE results, demonstrating that the modification made in the proposed GAPSCN is beneficial to enhance the performance.

\section{S-GAPSCN for QPS Compression and Reconstruction}

\par In this section, we proposed an asymmetric model S-GAPSCN whose architecture of the decoder is significantly simpler than the encoder to meet the practical lightweight design requirement at the IRS side. In other words, the computational complexity of the decoder is much lower than the encoder. Besides, to compensate for the performance degradation caused by simplifying the model’s architecture, we design a low computation complexity JAAMSN module in the decoder of S-GAPSCN. Moreover, to evaluate the effeteness of the JAAMSN module, we compare the performance between JAAMMSN and the AGSN.
\begin{figure*}[t]
\centering
\includegraphics[width=7in]{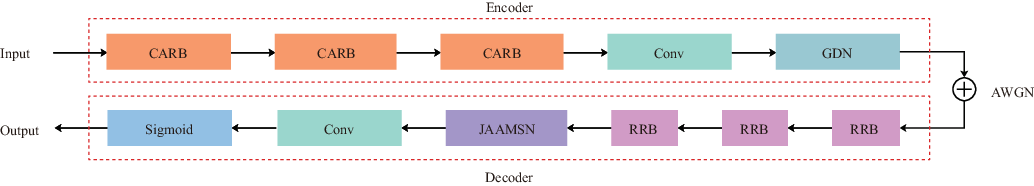}
\caption{The architectures of the S-GAPSCN.}
\label{SGAPSCN}
\end{figure*}

\begin{figure*}[t]
\centering
\includegraphics[width=7in]{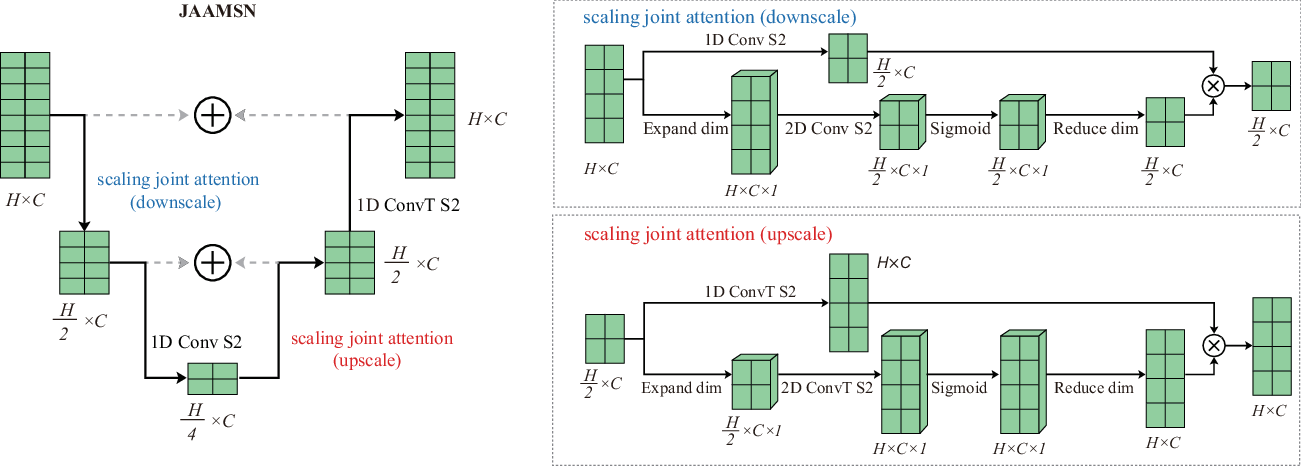}
\caption{The architectures of the S-GAPSCN, scaling joint attention (upscale) and scaling joint attention (downscale), where S2 in JAAMSN stands for the convolutional layer or transpose convolutional layer with stride 2.}
\label{JAAMSN}
\end{figure*}

\subsection{The Proposed S-GAPSCN}

\par Features scaling architectures (e.g., downscale and upscale) have achieved tremendous success in image processing, including the well-known U-Net\cite{Ronneberger} adopted for 2D biomedical image segmentation, three-dimensional (3D) image segmentation\cite{Fausto}, and image restoration\cite{LiuPengju, Park, LiuWei}. In addition, increasing attention has been paid to image denoising using feature scaling architectures such as \cite{JiaXixi, LBao, FJia, YZhang}. In contrast to the aforementioned works, an attention-guided scaling network (AGSN) for accurate image denoising was investigated in \cite{YZhang}, which employed channel attention and spatial attention in multi-scale operation to enhance the denoising performance. It should be noted that the above solutions are not suitable for S-GAPSCN since two attention processes are associated, which incur a high computational cost.

\par Inspired by the success of feature scaling architectures in image processing, we propose a lightweight JAAMSN to compensate for the performance degradation caused by simplifying the model parameter of the decoder, as shown in Fig.\ref{JAAMSN}. The details of JAAMSN are shown in Fig.\ref{JAAMSN}, which consist of two downscale processes (the first with attention mechanisms) and two upscale processes (the first with attention mechanisms). Similar to AGSN, we use a convolutional layer with stride 2 to downscale the data and a transpose convolutional layer with stride 2 to upscale the data. The attention downscale process consists of two branches. The first branch generates the attention map by adopting simplified global attention, which expands one dimension for features and employs a 2D convolutional layer with a nonlinear activation function sigmoid to produce the attention map and reduce the expanded dimension to obtain the final attention map. The second branch uses the convolutional layer to extract the most important information from input data. In the end, we construct an element-wise product between the outputs of the two branches. The only difference between the upscale and downscale processes is that we change the 2D convolutional layer to a 2D transpose convolutional layer for the upscale process, while there is no such change for the downscale process. The JAAMSN can be formulated as follows:
\begin{equation}
\begin{split}
    AM_D&=\sigma(f_{rd}(f_{2dConv}(f_{ep}(X))),\\
    ATT_D&=f_{2dConv}(X)\otimes AM_D,\\
    AM_U&=\sigma(f_{rd}(f_{2dTConv}(f_{ep}(X))),\\
    ATT_U&=f_{2dTConv}(X)\otimes AM_U,\\
    Y&=(f_{2dTConv}(ATT_U(f_{2dConv}(ATT_D(X)))\\
    &+ATT_D(X))+X),
\end{split}
\end{equation}

where $f_{2dTConv}$ denotes the 2D transpose convolutional layer. It should be noted that our proposed JAAMSN is significantly different from AGSN\cite{YZhang} through the scaling joint attention mechanism. The JAAMSN aims to emphasize meaningful features along the joint channel-spatial dimension and suppresses the AWGN effect by adopting a structure of multi-scale and scaling joint attention. Moreover, the proposed scaling joint attention in JAAMSN computes attention only once, instead of twice in channel-spatial attention, indicating a lower computational complexity in JAAMSN than that in AGSN.

\par As shown in Fig.~\ref{SGAPSCN}, S-GAPSCN consists of an encoder and a decoder. This encoder follows the same architecture as the encoder used in GAPSCN. By utilizing the RB as a first step in decoding, we are able to reconstruct the QPS and recover it to its original dimensions. Once the QPS data dimension has been reconstructed, the JAAMSN is used to suppress the AWGN.

\subsection{Comparison with AGSN}
\begin{table}[t]
\caption{Performance of MSE and Accuracy}
\centering
\begin{tabular}{|c|c|c|c|}
\hline
CR&Method &JAAMSN&AGSN\cite{YZhang}\\
\hline
\multirow{1}[2]*{$CR=\frac{1}{2}$}&Validation loss &$6.34\times10^{-6}$ &$5.97\times10^{-5}$ \\\cline{2-4} & Validation acc&0.99999 &0.99992\\
\hline
\multirow{1}[2]*{$CR=\frac{1}{4}$} &Validation loss  &0.0446 &0.0513\\\cline{2-4}   & Validation acc &0.9369 &0.9316\\
\hline
\multirow{1}[2]*{$CR=\frac{1}{8}$} &Validation loss &0.1535 &0.1628\\\cline{2-4} & Validation acc &0.7611 &0.7604\\
\hline
\end{tabular}
\label{sga}
\end{table}

\par In order to assess the relative effectiveness of different attention-guided network modules, we select the S-GAPSCN architecture as the basic architecture and train and test it by switching between JAAMSN and AGSN. As with the comparisons in Section IV, we chose MSE and accuracy functions as the criteria to evaluate model performance. For ease of illustration, we select the validation loss and validation accuracy when the training process converges. The validation loss and validation accuracy performance are presented in Table~\ref{sga}. Based on Table~\ref{sga}, JAAMSN achieves more accurate results under each CR when compared with AGSN. The proposed scaling joint attention in JAAMSN is also capable of emphasizing meaningful features along the channel dimension and the spatial dimension, which is different from channel-spatial attention in AGSN, in a joint manner. In addition, the multi-scale architecture employed in the JAAMSN can alleviate the noise effect.

\par The channel-spatial attention in AGSN can emphasize meaningful features along the channel dimension and spatial dimension separately. Thus, the channel-spatial attention in AGSN computes attention maps twice, e.g., channel attention map and spatial attention map, while the scaling joint attention in JAAMSN only calculates the attention map once, e.g., the joint channel-spatial attention map. This implies that JAAMSN has a lower computational complexity than AGSN. Based on the above observations, JAAMSN is expected to achieve reliable reconstruction accuracy and low computational complexity, which will be evaluated and verified in Section VI.

\begin{table}[t]
\caption{Hyperparameter Settings}
    \centering
    \scalebox{1.2}{
    \begin{tabular}{|c|c|}
    \hline
    Neurons & 64\\
    \hline
    Batch size & 256 \\ 
    \hline
    Training epoch & 1000\\ 
    \hline
    Learning rate& $10^{-4}$\\
    \hline
    Optimizer & Adam\\
    \hline
    Loss function & Mean squared error\\
    \hline
    \end{tabular}}
    \label{parameter}
\end{table}

\section{Experimental Studies}

\subsection{Training Procedure}
In the simulations, an IRS-assisted downlink system consists of an IRS with $M=16$ reflecting elements is considered, as defined in Fig. 1. Each phase shift will be quantized to $K=8$ bits. The noise will be added to the output of the encoder in the online training. With the empirical results in \cite{J. Guo}, the training SNR we chose is 20dB. We set training epochs as 1000. We generate training samples, validation samples, and test samples according to the uniform distribution. The sizes of training samples, validation samples, and test samples are 128000, 32000 and 128000, respectively. We chose the Adam algorithm\cite{adam} as the optimizer to update the parameters and the learning rate is $10^{-4}$. The hyperparameter settings are summarized in Table.~\ref{parameter}.

\begin{figure}[t]
\centering
\includegraphics[width=3.8in]{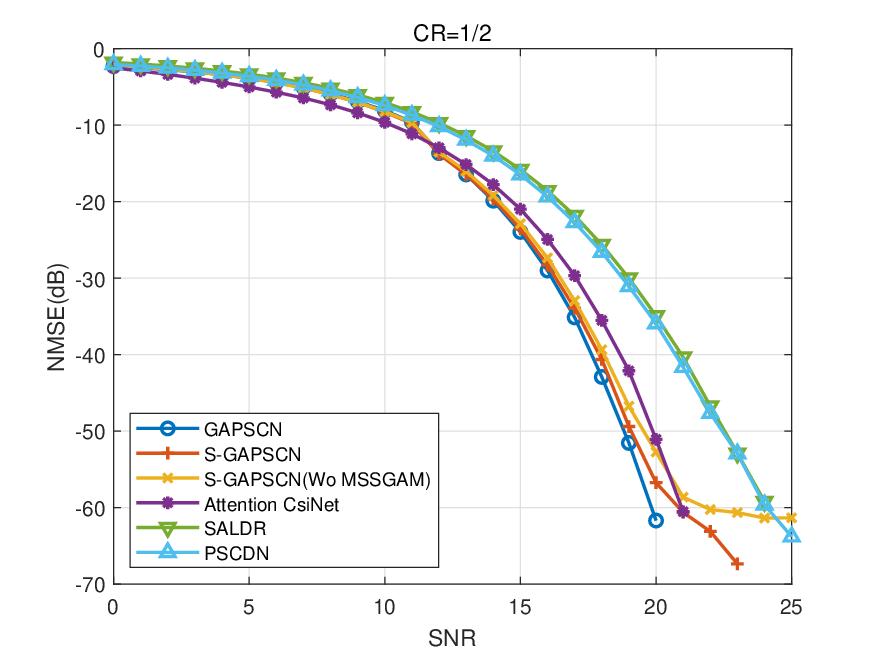}
\caption{NMSE performance (dB) comparison for different compression algorithms  under CR$=\frac{1}{2}$.}
\label{cr=1/2}
\end{figure}

\begin{figure}[t]
\centering
\includegraphics[width=3.8in]{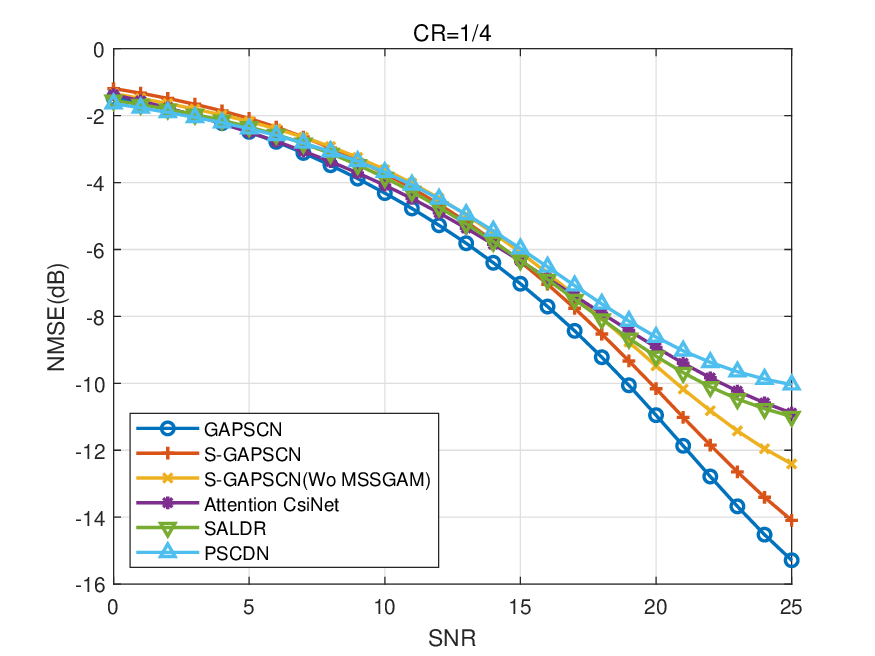}
\caption{NMSE performance (dB) comparison for different compression algorithms under CR$=\frac{1}{4}$.}
\label{cr=1/4}
\end{figure}

\begin{figure}[t]
\centering
\includegraphics[width=3.8in]{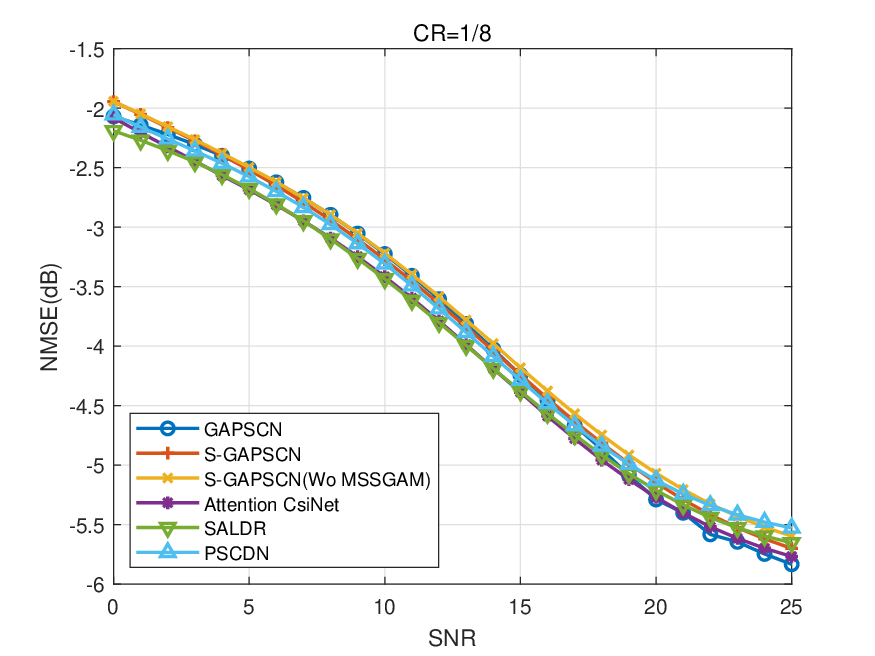}
\caption{NMSE performance (dB) comparison for different compression algorithms under CR$=\frac{1}{8}$.}
\label{cr=1/8}
\end{figure}
\begin{table*}[t]
\caption{Comparison of computational complexity and model size}
\centering
\begin{tabular}{|c|c|c|c|c|c|c|}
\hline
\multirow{2}{*}{Model}&\multicolumn{3}{|c|}{\textbf{Parameter}} &\multicolumn{3}{|c|}{\textbf{Processing time}} \\
\cline{2-7}
 &Encoder &Decoder &Total & Encoder &Decoder &Total\\
\hline
GAPSCN &274551 &290791 &565342 &2.05ms &2.35ms &4.40ms\\
\hline
S-GAPSCN &274551 &91157 &365708 &2.04ms &0.75ms &2.79ms\\
\hline
Attention CsiNet &4636 &2148670 &2153306 &0.31ms &1.55ms &1.86ms\\
\hline
SALDR &161808 &254660 &416468 &0.92ms &2.47ms &3.39ms\\
\hline
PSCDN &156048 &197121 &353169 &0.39ms &1.28ms &1.67ms\\
\hline
\end{tabular}
\label{tab:my_label}
\end{table*}

\subsection{Simulation Results and Analysis}

\par The NMSE is adopted as a criterion to demonstrate the prominent performance of the proposed GAPSCN and S-GAPSCN, which can be formulated as:
\begin{equation}
    {\rm NMSE} =\frac{\|X-\hat{X}\|^2_2}{\|X\|^2_2}.
\end{equation}

\par For comparison, we consider the following convolutional attention-based compression methods, including Attention CsiNet\cite{QCai} and SALDR\cite{XSong}, for performance comparison. The 2D convolutional layer in Attention CsiNet and SALDR is switched to a 1D convolutional layer since our data type has only one dimension, and other hyperparameter settings are kept the same as our proposed model. This serves as an ablation study of the effect of JAAMSN. In addition, we evaluate the model size and model complexity by calculating the number of parameters and examining the run time for the proposed S-GAPSCN.
 
\par The NMSE performance of GAPSCN, S-GAPSCN, S-GAPSCN without MASSGN, Attention Csineti, SALDR, and PSCDN with CRs $\frac{1}{2},\frac{1}{4},\frac{1}{8}$ is shown in Fig.\ref{cr=1/2}, Fig.\ref{cr=1/4}, and Fig.\ref{cr=1/8}, respectively. In Fig.\ref{cr=1/2}, Fig.\ref{cr=1/4}, and Fig.\ref{cr=1/8}, the performance of all compression methods improves with increasing CR, as expected. Furthermore, GAPSCN consistently achieves superior NMSE performance compared to the two benchmark schemes for each CR, which can be attributed to the following modifications in GAPSCN. First, by incorporating the missing correlated information between spatial and channel dimensions in the global attention, a more comprehensive attention map is established, capturing more relevant information. Second, we abandon the pooling operation to fully utilize the information. Finally, we rearrange the data distribution throughout the model using the GND/IGDN layer. The JAAMSN significantly enhances S-GAPSCN's performance, as demonstrated in the ablation study results, which show S-GAPSCN outperforms S-GAPSCN without JAAMSN, and a small gap exists between S-GAPSCN and GAPSCN. This result is due to the fact that JAAMSN can capture critical information and mitigate the noise effect by employing a simplified global attention and multi-scale architecture.

\par In Table III, we compare the model size and computational complexity for different compression methods under a CR of $\frac{1}{8}$. According to the results, the decoder of S-GAPSCN has the fewest parameters and the fastest processing time. For example, compared with another lightweight design model, PSCDN, the decoder of S-GAPSCN requires only 0.75ms, while PSCDN needs 1.28ms, resulting in a $42.4\%$ improvement. Although the total computational complexity of S-GAPSCN increases, it is the most suitable for IRS-assisted wireless systems, as the IRS is the only component in these systems that faces a shortage of computational resources. The aforementioned results demonstrate that S-GAPSCN achieves a prominent NMSE performance while maintaining low computational complexity.

\section{Conclusion}
In this paper, we investigated and analyzed the problem of QPS signaling overhead in IRS-assisted wireless systems by proposing a novel global attention-based model, GAPSCN, and a low computational complexity model, S-GAPSCN. In the proposed GAPSCN, we first supplement the missing correlated information between spatial and channel dimensions in the global attention, resulting in a more comprehensive attention map that allows the model to emphasize a greater amount of meaningful information. We further alleviate the AWGN effect by utilizing the GND/IGDN layer. Additionally, we propose S-GAPSCN to reduce the decoder complexity for the IRS, where the JAAMSN captures important information and mitigates the noise effect by employing a joint attention scheme and multi-scale architecture, respectively. Simulation results show that the proposed GAPSCN is capable of achieving accurate reconstruction accuracy compared with existing state-of-the-art models in wireless communications, while S-GAPSCN can provide nearly equivalent performance but at a much lower computational cost compared to GAPSCN.





\vfill

\end{document}